\documentstyle{jjaptex}
\def\nel{N_{\rm el}}

\def\beeq{\begin{equation}}
\def\eneq{\end{equation}}
\def\beeqa{\begin{eqnarray}}
\def\eneqa{\end{eqnarray}}
\def\theequation{\arabic{equation}}

\def\thesection{\Roman{section}.}
\def\thepage{-- \arabic{page} --}
\title{Nonlinear Optical Response from Excitons\\
in Soliton Lattice Systems of Doped Conducting Polymers}
\author{Kikuo H\sc{arigaya}\footnote{E-mail address: 
harigaya@etl.go.jp; URL: http://www.etl.go.jp/People/harigaya/.}}
\inst{Physical Science Division, Electrotechnical Laboratory,
Umezono 1-1-4, Tsukuba, Ibaraki 305, Japan}
\recdate{February 14, 1997}
\accdate{March 21, 1997}
\abst{%
Exciton effects on conjugated polymers and the resulting 
optical nonlinearities are investigated in soliton lattice 
states.  We use the Su-Schrieffer-Heeger model with long-range 
Coulomb interactions for electrons.  The third-harmonic generation (THG) 
at off-resonant frequencies is calculated as functions of 
the soliton concentration and the chain length of the polymer.  
The magnitude of the THG with 10 percent doping increases 
by the factor of about 10$^2$ from that of the neutral system.  
The large increase of the order two is common for several 
Coulomb interaction strengths, and is seen in open
systems as well as in periodic systems.}
\kword{excitons, nonlinear optical response,
soliton lattice systems, doped conducting polymers, theory}
\begin{document}
\maketitle
\sloppy
The doping effects in conjugated polymers and their linear and 
nonlinear optical responses are fascinating research topics 
because of their importance in scientific interests as well 
as in developing technology.  In the previous paper,\cite{1} 
we theoretically considered exciton effects in the soliton 
lattice states of doped systems.  There is one type of exciton 
in the undoped system with half-filled electronic states,\cite{2}
where the excited electron (hole) sits at the bottom of the 
conduction band (top of the valence band), which we called 
the intercontinuum exciton.\cite{1}  In the 
soliton lattice states of the doped Su-Schrieffer-Heeger 
(SSH) model for degenerate conjugated polymers,\cite{3} there are 
small energy gaps between the soliton band and the continuum states, 
i.e., valence and conduction bands.  The number of the types 
of excitons increases, and their presence effects structures 
of the optical spectra.  A new exciton, which we have named the 
soliton-continuum exciton,\cite{1} appears when the electron-hole 
excitation is considered between the soliton band and one of 
the continuum bands.

We are conducting this study to investigate how large optical 
nonlinearities are obtained when conjugated polymers are doped with 
electrons or holes up to as much as 10 percent.  The SSH model,\cite{3} 
with the long range Coulomb interactions of the Ohno expression,\cite{4} 
is solved with the Hartree-Fock (HF) approximation, and 
the excitation wavefunctions of electron-hole pairs are
calculated by the single-excitation configuration-interaction 
(single-CI) method.  We calculated the off-resonant nonlinear 
susceptibility as a guideline of the magnitude of the nonlinearity.  
We looked at the characters of the excitations at off-resonances 
resulting from the single-CI calculations.  The third harmonic 
generation (THG) at the zero 
frequency, $\chi_{\rm THG}^{(3)} (0) = \chi^{(3)} 
(3\omega;\omega,\omega,\omega)|_{\omega = 0}$, is 
considered with changing of the chain length and the soliton 
concentration.  We will demonstrate that the magnitude of the THG 
at the 10 percent doping increases by the factor of about 10$^2$ 
from that of the neutral system.

The periodic systems simulate the reasonably long systems where chain end
effects are small.  However, all of the polymers are not 
sufficiently long enough to neglect end effects, and shorter 
chains are present.  The chain ends might have effects on 
the electronic structures, excitation properties and 
the resulting nonlinear optical responses of the polymers.  In this 
paper, we discuss how the chain ends effect the
large increase of optical nonlinearity, compared with that of
the periodic systems.  We will demonstrate that the large 
increase of the THG upon doping is seen in open systems
as well as in periodic systems.

We use the SSH hamiltonian\cite{3} with the Coulomb interactions,
\beeq
H = H_{\rm SSH} + H_{\rm int}.
\eneq
The first term of eq. (1) is,
\beeqa
H_{\rm SSH} &=& - \sum_{i,\sigma} ( t - \alpha y_i )
( c_{i,\sigma}^\dagger c_{i+1,\sigma} + {\rm h.c.} )  \nonumber \\
&+& \frac{K}{2} \sum_i y_i^2,
\eneqa
where $t$ is the hopping integral of the system without the 
bond alternation; $\alpha$ is the electron-phonon coupling 
constant which changes the hopping integral linearly with 
respect to the bond variable $y_i$; the bond variable is assigned
to the bond between the $i$th and $(i+1)$th sites ($1 \leq i
\leq N -1$, $N$ being the system size); $c_{i,\sigma}$ is an 
annihilation operator of the $\pi$-electron at the site $i$ 
with spin $\sigma$; the sum is taken over $i$ within $1 \leq i
\leq N -1$ due to the open boundary conditions; and the 
last term with the spring constant $K$ is the harmonic energy of 
the classical spring simulating the $\sigma$-bond effects.  The 
second term of eq. (1) is the long-range Coulomb interaction in 
the form of the Ohno potential:\cite{4}
\beeqa
H_{\rm int} &=& U \sum_i 
(c_{i,\uparrow}^\dagger c_{i,\uparrow} - \frac{n_{\rm el}}{2})
(c_{i,\downarrow}^\dagger c_{i,\downarrow} - \frac{n_{\rm el}}{2}) \nonumber \\
&+& \sum_{i \neq j} W(r_{i,j}) 
(\sum_\sigma c_{i,\sigma}^\dagger c_{i,\sigma} - n_{\rm el})
(\sum_\tau c_{j,\tau}^\dagger c_{j,\tau} - n_{\rm el}),
\eneqa
where $n_{\rm el}$ is the number of $\pi$-electrons per site,
$r_{i,j}$ is the distance between the $i$th and $j$th sites, and
\beeq
W(r) = \frac{1}{\sqrt{(1/U)^2 + (r/a V)^2}}
\eneq
is the Ohno potential.  The quantity $W(0) = U$ is the strength of 
the onsite interaction, $V$ is the strength of the long range part,
and $a$ is the mean bond length.

The model is treated by the HF approximation and the single-CI 
for the Coulomb potential.  The bond variables are calculated 
by the adiabatic approximation.  The selfconsistent formalism 
was explained in our previous paper.\cite{1}  The electric 
field of light is parallel to the polymer chain which is along 
the $x$-axis.  The optical absorption spectra are calculated by 
the formula
\beeq
\sum_\kappa E_{\kappa} P (\omega - E_{\kappa}) 
\langle g | x |\kappa \rangle \langle \kappa | x | g \rangle,
\eneq
where $P (\omega) = \gamma/[ \pi (\omega^2 + \gamma^2)]$
is the Lorentzian distribution ($\gamma$ is the width),
$E_{\kappa}$ is the electron-hole excitation energy, 
$| \kappa \rangle$ is the $\kappa$th excitation, and
$| g \rangle$ is the ground state.

The THG is calculated with the conventional formula,\cite{5,6,7}
which is sometimes called the sum-over-states method.  
In order to demonstrate the magnitude of the 
THG, we use the number density of the CH unit, which is 
taken from {\sl trans}-polyacetylene: $5.24\times 10^{22} 
{\rm cm}^{-3}$.\cite{8}  We also use $t=1.8$ eV in order to look at 
numerical data in the esu unit.  We include a small 
imaginary part $\eta$ in the denominator, which assumes a 
lifetime broadening and suppresses the height of the 
$\delta$-function peaks.  The THG at $\omega = 0$ does not
sensitively depend on the choice of $\eta$.  This can be
checked by varying the broadening.  Here, we report the 
results with the value $\eta = 0.02 t$.

The system size is chosen as $N= 80$, 100, 120 when the 
electron number is even (it is varied from $\nel = N, N+2, 
N+4, N+6, N+8, N+10$ to $N+12$).   We change Coulomb 
interaction parameters arbitrarily within a reasonable range in 
order to look at general properties of the optical 
nonlinearities of the soliton lattice systems.  We take two 
combinations of the Coulomb parameters $(U,V) = (2t,1t)$ and 
$(4t,2t)$ as representative cases.  The other parameters, 
$t = 1.8$ eV, $K = 21$ eV/\AA$^2$ and $\alpha = 4.1$ eV/\AA, 
are fixed in view of the general interests of this study.  
All the quantities of the energy dimensions are shown in 
units of $t$.

Figure 1 shows the typical lattice configuration and 
excess-electron density distribution of open systems for 
$N=100$, $\nel = 102$, $U=4t$ and $V=2t$.  Both quantities are 
smoothed by removing small oscillations between even- 
and odd-number sites.  There are two charged solitons 
due to the excess-electron number $\nel - N = 2$.  The 
two solitons are centered around the 20th and 80th 
lattice sites.  They would be around the 25th and 75th 
sites in the periodic boundary system.  The solitons are 
moved slightly closer to the chain ends.  This is the end 
effect, due to the enhanced bond-alternation strengths 
near the 1st and 100th sites.  The enhanced bond variables 
pull the two solitons in the direction of the chain ends.  
These effects have already been seen in an earlier paper.\cite{9}
The excess-electron density shows that the doped charges 
accumulate with the maxima at the soliton centers.

Following, we calculate the optical spectra, linear 
absorption and THG, and consider the exciton effects.  Figure 2 
(a) shows the typical optical absorption spectrum at the 2 \% 
soliton concentration for $(U,V) = (4t,2t)$.  The broadening 
$\gamma = 0.05t$ is used.  There are two main features around 
the energies 0.7$t$ and 1.4$t$.  The former originates from 
the soliton-continuum exciton, and the latter is from the 
intercontinuum exciton.   The presence of the chain ends does 
not effect the excitation energies so much.  However, the 
oscillator strengths of the soliton-continuum exciton become 
relatively larger than those in the periodic system.  This is a new 
property which is found in the system with open boundaries.
We have treated the periodic system with ring geometries,
and the open system with geometries of linear chains.  Thus,
the expectation values of dipole operators are generally larger 
in the open system than in the periodic system.  A similar 
fact has been discussed for the uniformly dimerized system 
without solitons in ref. \citen{10}.

Figure 2 (b) displays the absolute value of the THG against 
the excitation energy $\omega$.  The abscissa is scaled by a 
factor of 3 so that the features in the THG are located at similar 
points to those in the abscissa of Fig. 2 (a).  The large feature at 
approximately $\omega=0.22t$ originates from the lowest excitation of the 
soliton-continuum exciton and the other feature at approximately 
$\omega=0.26t$ originates from the higher excitations.  The features 
from the intercontinuum exciton extend from $\omega = 0.48t$ 
to the higher energies.  The soliton-continuum exciton gives 
rise to the large optical nonlinearity, as we looked in the 
linear absorption.  The oscillator strengths accumulate at the 
lowest excited state of the soliton-continuum exciton while
the doping proceeds, as we have shown in ref. \citen{1}.  This feature
mainly contributes to the large optical nonlinearity of the
doped systems.  The THG spectra, such as in Fig. 2 (b), are 
calculated for the three system sizes, $N=80$, 100, 120, 
and for the soliton concentrations up to 10 \%, in order 
to compare to the calculations for the system with periodic boundaries.

Figures 3 (a) and 3 (b) display the variations of the absolute
value of $\chi_{\rm THG}^{(3)} (0)$ for $(U,V) = (2t,1t)$ 
and $(4t,2t)$, respectively.  The plots are the numerical 
data: the open symbols are for the periodic systems, 
and the closed symbols are for the present 
calculations of the systems with chain ends.  The dashed 
lines are guides for the plots of the periodic 
systems, showing the overall behavior for each system size.  
The finite system size effects\cite{11} appear in these figures.  It is 
known that the THG is not size consistent, and spectral 
shapes depend on the system size when $N$ is as large as 
100.\cite{11}  This fact is reflected in the separation
of the plots of the three system sizes with periodic and 
open boundaries.

The off-resonant THG near zero concentration increases very 
rapidly, but the THG still increases for a few percent up to 
a 10 \% soliton concentration.  The increase between the zero 
concentration and the 10 \% concentration is by the factor 
of approximately 100.  This behavior is consistent for the two boundary 
conditions and the Coulomb interaction parameters.  The main 
difference between the two boundary conditions is that the 
THG with open boundaries is larger than that of the periodic 
system, when the concentration, system size, and the Coulomb 
parameters are the same.  A similar property has been 
discussed in the calculations of the half-filled systems.\cite{10}

When we look at the enhancement factor in detail, it is 
discovered that the enhancement is particularly large for 
the open boundary systems with two solitons.
This is seen for the three closed symbols near
2 \% concentration in Figs. 3 (a) and 3 (b).  These are 
the cases of the concentrations of 2.5 \%, 2.0 \%, and 1.67 \%,
for $N=80$, 100, and 120, respectively.  Therefore, if the
limit of the long chain is taken, there is a jump of the THG
between the zero concentration and the infinitesimal concentration.
This is different from the expected smooth behaviors of the
periodic chains.  The presence of the chain ends results in
the trapping of solitons near the ends, and thus the THG
jumps suddenly at the zero concentration limit.

The dramatic enhancement of the THG in the case of two solitons 
is clear when we consider the ratio of the THG value of the 
open system with respect to that of the periodic system.  The 
calculations have been done for $(U,V)=(4t,2t)$, and the ratio 
is shown in Fig. 4.  The numerical data are shown by 
the triangles ($N=80$), circles ($N=100$), and squares ($N=120$).
Except for the three plots near the concentration of 2 \%, 
the ratio is constant at approximately 3.  It becomes greater
than 10 for the two soliton systems.  Thus, we have found that
the case with two solitons is special, mainly due to the
chain end effects.  It should be noted that the large THG
due to the presence of chain ends could be used as a tool
for increasing nonlinear optical responses experimentally.
The reason for the large enhancement can be understood as follows.
In two soliton solutions with the open boundary condition, 
the intersoliton distance becomes longer than in the system
with periodic boundaries.  This gives rise to the larger 
expectation values of the dipole moment operators, and thus 
we obtain the greatly enhanced THG in the system with two 
solitons.

In summary, we have considered the off-resonant nonlinear 
susceptibility as a guideline of the strength of the 
nonlinearity in the doped conjugated polymers.   We have 
calculated the off-resonant THG with various system
sizes and the soliton concentration for the chains with open 
boundaries.  We have shown that the magnitude of the THG at 
10 percent doping increases by the factor of 10$^2$ 
from that of the neutral system.  The large increase of the 
order two is common for several Coulomb interaction strengths, 
and is seen in the open systems 
as well as in the periodic systems.

\mbox{}

\noindent
{\bf Acknowledgments}\\
The author acknowledges valuable discussions with 
Professor T. Kobayashi, Professor S. Stafstr\"{o}m, Dr. S. Abe, 
Dr. Y. Shimoi, and Dr. A. Takahashi.

\begin{halffigure}
\caption{(a)  The bond alternation order parameter, 
$(-1)^n (y_{n+1} - y_n)/2$, and (b) the excess-electron 
density, $(\rho_{n-1} + 2 \rho_n + \rho_{n+1})/4$.  
The parameters are $U=4t$, $V=2t$, $N=100$, and 
$\nel = 102$.  See the text for the other parameters.}
\label{Fig:1}
\end{halffigure}

\begin{halffigure}
\caption{(a) The optical absorption spectrum and (b) the absolute
value of the THG, for the open system with the size $N=100$, the 
electron number $\nel = 102$, and $(U,V) = (4t,2t)$.  The broadening
$\gamma = 0.05t$ is used in (a), and $\eta = 0.02t$ is used
in (b).  The absorption is shown in arbitrary units,
and the nonlinear optical response is in esu units.}
\label{Fig:2}
\end{halffigure}

\begin{halffigure}
\caption{The absolute value of the THG at $\omega = 0$ v.s. 
the soliton concentration for (a) $(U,V) = (2t,1t)$ and 
(b) $(4t,2t)$.  The numerical data are shown by the triangles 
($N=80$), circles ($N=100$), and squares ($N=120$).
The data of the system with the periodic boundary condition 
are shown by the open symbols.  And, the data with the open 
boundaries are shown by the closed symbols.  The dashed lines 
are guides.}
\label{Fig:3}
\end{halffigure}

\begin{halffigure}
\caption{The ratio of the absolute values of the THG
between the open and periodic boundary conditions, 
shown against the soliton concentration.  The Coulomb 
parameters are $(U,V) = (4t,2t)$.  The numerical data 
are shown by the triangles ($N=80$), circles ($N=100$), 
and squares ($N=120$).}
\label{Fig:4}
\end{halffigure}

\makefigurecaptions
\end{document}